\documentclass[preprint,superscriptaddress, aps, prb]{revtex4-2}
\usepackage{amsmath}
\usepackage{xcolor}
\usepackage{graphicx}
\usepackage{mathptmx}
\usepackage{bm}
\usepackage[colorlinks=true,urlcolor=blue,linkcolor=blue,citecolor=blue]{hyperref}
\newcommand{\tstern}{\ensuremath{T_2^*} }

\newcommand*{\fg}[1]{Fig.\thinspace\ref{#1}}

\begin{document}

\title{Trembling motion of electrons driven by Larmor spin precession}

\author{I. Stepanov}
\affiliation{2nd Institute of Physics and JARA-FIT, RWTH Aachen University, D-52074 Aachen, Germany}

\author{M. Ersfeld}
\affiliation{2nd Institute of Physics and JARA-FIT, RWTH Aachen University, D-52074 Aachen, Germany}

\author{A. V. Poshakinskiy}
\affiliation{Ioffe Institute, 194021 St Petersburg, Russia}

\author{M. Lepsa}
\affiliation{Peter Gr\"{u}nberg Institut (PGI-9) and JARA-FIT, Forschungszentrum J\"{u}lich GmbH, D-52425 J\"{u}lich, Germany}

\author{E. L. Ivchenko}
\affiliation{Ioffe Institute, 194021 St Petersburg, Russia}

\author{S. A. Tarasenko}
\affiliation{Ioffe Institute, 194021 St Petersburg, Russia}

\author{B. Beschoten}
\thanks{e-mail: bernd.beschoten@physik.rwth-aachen.de}
\affiliation{2nd Institute of Physics and JARA-FIT, RWTH Aachen University, D-52074 Aachen, Germany}

\begin{abstract}
We show that the initialization of an ensemble of electrons in the same spin state in strained $n$-InGaAs subject to a perpendicular magnetic field triggers an AC electric current at GHz frequencies. The AC current emerges in the absence of any driving force and survives until the coherent precession of the electron spins is lost. The current amplitude increases linearly with both the spin-orbit coupling strength and the external magnetic field. The generation mechanism of the observed oscillatory charge motion can be fruitfully described in terms of the periodic trembling motion of spin-polarized electrons, which is a solid-state analog to the Zitterbewegung of free Dirac electrons. Our results demonstrate that the hidden consequence of relativistic quantum mechanics is realized and can be studied in a rather simple solid-state system at moderate temperatures. Furthermore, the large amplitude of the AC current at high magnetic fields enables ultra-fast spin sensitive electric read-out in solids.
\end{abstract}

\maketitle
 \newpage
The search for efficient and fast techniques to generate, manipulate and detect spin polarized carriers in semiconductors by pure electrical means is at the core of modern solid-state spintronics.
%~\cite{Awschalom2007Mar,Fabian2007Aug}. 
It has been shown that these tasks can be fulfilled in III-V semiconductors by the use of spin-orbit interaction (SOI) providing a rich platform to couple the charge and spin degree of freedom~\cite{Awschalom2009Jun}. SOI can convert a charge current into a homogeneous spin polarization 
(current-induced spin polarization)~\cite{IvchenkoGanichev2017,Kato2004Oct,Silov_2004,Ganichev2005,Stern2006Sep}
%\cite{Edelstein1990Jan,Kato2004Oct,Sih2005Sep,Stern2006Sep} 
or into spin currents transverse to the charge current direction by spin dependent scattering (spin Hall effect)~\cite{Dyakonov2017,Kato1910,Wunderlich2005Feb,Stern2008Sep}.
%\cite{Hirsch1999Aug,Zhang2000Jul,Kato1910,Wunderlich2005Feb,Sih2005Sep,Stern2008Sep}. 
SOI can be also used for spin manipulation as it acts as an effective magnetic field where the spin precession can be controlled either by static or pulsed electrical field~\cite{KalevichKorenev1990,Kato2004_manipulation,Meier_Rashba_2007,apl1.4864468,PhysRevLett.109.146603}. For electrical spin detection, spin-polarized electrons can drive a charge current by either the spin-galvanic effect~\cite{Nature417_Ganichev2002_Spin-GalvanicEffect} 
or by the inverse spin Hall effect~\cite{Werake2011Mar}. While most of these experiments have been extended into the time-domain by combining ultrafast electrical and optical techniques~\cite{Kato2004Oct,Stern2008Sep,PhysRevLett.109.146603,apl1.4864468,Schmidt_APL2015,Schreiber2021Nov}, there have not been any studies on spin sensitive ultrafast electrical readout.

We show that electrons in solids can themselves drive an AC electric current if initialized in the same spin state. The effect is demonstrated for semiconductor epilayers subject to an in-plane static magnetic field. The AC current emerges in the absence of any driving force and is maintained until the spin ensemble dephases. The observed phenomenon can be described in a semi-classical way as an electric current caused by the spin precession in structures with spin-orbit coupling~\cite{Ivchenko_JETP_1990}. Alternatively, it can be viewed in a quantum-mechanical picture as a trembling motion of electrons, a phenomenon similar to the Zitterbewegung of Dirac electrons~\cite{Original_Zitterbewegung,Zawadzki2011Mar,Schliemann_PhysRevLett.94.206801,Biswas_2012, Winkler_PhysRevB.75.205314,Manchon.2015}. The oscillating contribution to the electron velocity originates from the interference of the spin states separated by the Zeeman gap. The AC electric current is observed together with the spin-galvanic effect contribution and can be distinguished by phase and magnetic field-dependence.

\begin{figure*}[tbp]
\includegraphics[width=\linewidth]{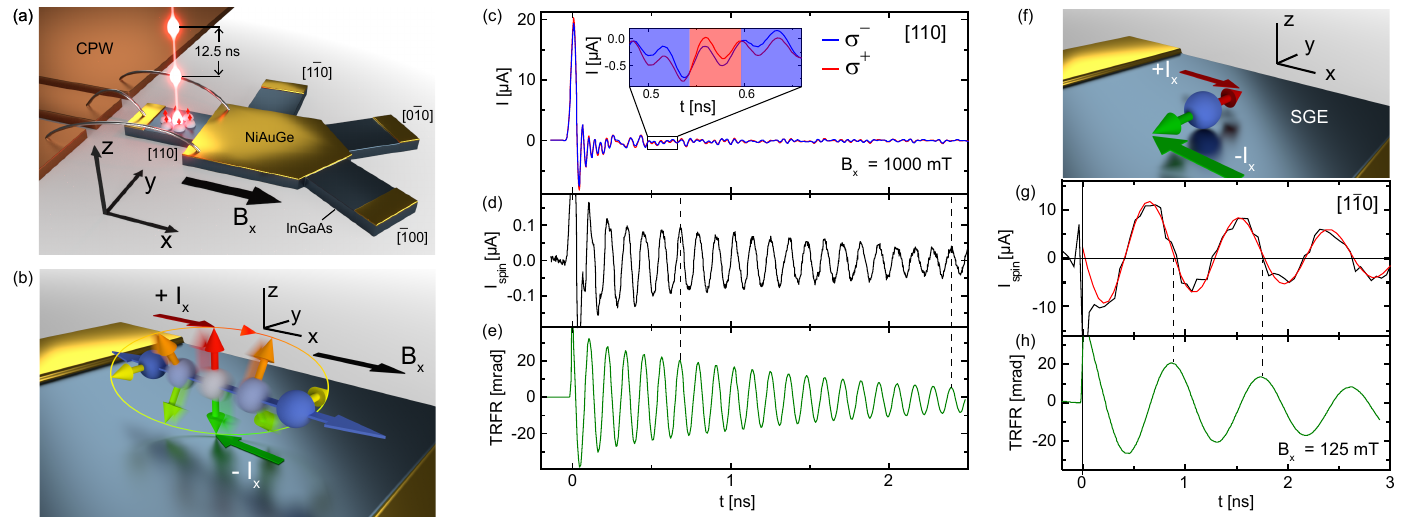}
\caption{\label{fig1} (a) Sample geometry and pulsed optical spin excitation. The electron spin ensemble is excited along the $z$ direction by circularly polarized ps laser pulses with a pulse repetition time of 12.5 ns. The high-frequency current in InGaAs is detected by a phase-triggered sampling oscilloscope through coplanar waveguides (CPW). Four different crystal directions can be contacted. (b) Illustration of the spin precession driven trembling electron motion (PDTM).
Electron spin precession yields a periodic displacement of the electrons along the $x$-direction which results in an AC current oscillating at the Larmor frequency. (c) Time-resolved current traces along the $[110]$ crystal direction recorded after pulsed optical excitation for $\sigma^+$ and $\sigma^-$ polarizations at $B_x = 1$\,T and $T = 50$~K and averaged over about $10^5$ measurements. The inset shows a close-up of both current traces. (d) Time-dependent AC current $I_{\rm spin} = [I(\sigma^+)-I(\sigma^-)]/2$ as determined from both traces in panel (c).  TRFR of electron spin precession under identical experimental conditions. (f)  Illustration of AC current generated by the spin-galvanic effect (SGE), where the largest currents are generated for electron spins pointing along the $\pm$ y-directions. (g) and (h) show respective AC current and TRFR traces measured along the $[1\overline{1}0]$.}
\end{figure*}

The experiments were performed on $n$-type In$_{0.07}$Ga$_{0.93}$As (\mbox{$n\sim 8.25\times 10^{16}$~cm$^{-3}$}) samples along different crystal axes (see Methods Section and \fg{fig1}(a) for the experimental geometry). Circularly polarized ps laser pulses with photon energies of 1.41~eV near the fundamental band edge of InGaAs are used to initiate the electron spin polarization $S_z(0)$ at $t=0$~ns pointing along $+z$ and $-z$ directions for $\sigma^+$ and $\sigma^-$ polarizations, respectively~\cite{Meier_Optical_Orientation}. The sample is subject to an external static magnetic field $\bm B~\parallel~x$. The generated AC electric current is probed through high frequency contacts by a sampling oscilloscope~\cite{PhysRevLett.109.146603,apl1.4864468}. 

After optical excitation of an unbiased sample by circularly polarized pulses we detect an AC electric current at GHz frequencies. The time-resolved current signal measured at $T=50$~K along the $\bm B$ field direction ($B_x=1$\,T) parallel to the [110] crystal axis is shown in \fg{fig1}(c) for $\sigma^+$ and $\sigma^-$ excitation. For short $t$, it is dominated by a spin-independent background which decays on the electron-hole recombination time of about 100 ps and is followed by some circuit ringing due to internal reflections in the transmission line. A spin-polarization-driven AC current is already visible in this data as subtle periodic differences between both curves which is shown in the close-up region in \fg{fig1}(c). We can enhance the visibility of the spin-dependent signal by plotting $I_{\rm spin}=[I(\sigma^+)-I(\sigma^-)]/2$ in \fg{fig1}(d). An oscillating current is clearly visible. It extends over several ns after the laser-induced non-equilibrium electron-hole population has recombined indicating that the spin angular momentum has been transferred to the resident electron ensemble. The current oscillations follow the Larmor precession of the electron spins as seen from time-resolved pump-probe Faraday rotation (TRFR) data in \fg{fig1}(e) measured under identical experimental conditions. Furthermore, both traces follow the same temporal decay showing that the same spin states are probed by both techniques. While Larmor spin precession of optically generated spin packets has routinely been probed by time-resolved optical techniques~\cite{Kikkawa1998,Heberle1994Jun,Sih2006Jun,Greilich2006Jul}, it has never been measured in the time-domain by electrical means.

The AC current trace in \fg{fig1}(d) can be fitted by an exponentially decaying cosine function
\begin{equation}
I_{\rm spin}(t) = I_0 \exp\left(-\frac{t}{\tstern}\right) \cos\left(\omega_L t+\phi\right),
\label{Oszillation}
\end{equation}
to extract the current amplitude $I_0$, the spin dephasing time $\tstern$ and the respective phase $\phi$, with the Larmor frequency
$\omega_L$ determined by the electron $g$-factor $|g|=0.62$. We note that the AC current in \fg{fig1}(d) is in phase ($\phi = 0^{\circ}$) with the $z$-component of the spin polarization which is directly probed by TRFR (for comparison of the relative phases see dotted lines in \fg{fig1}(d) and (e)).

\begin{figure}[tb]
\includegraphics{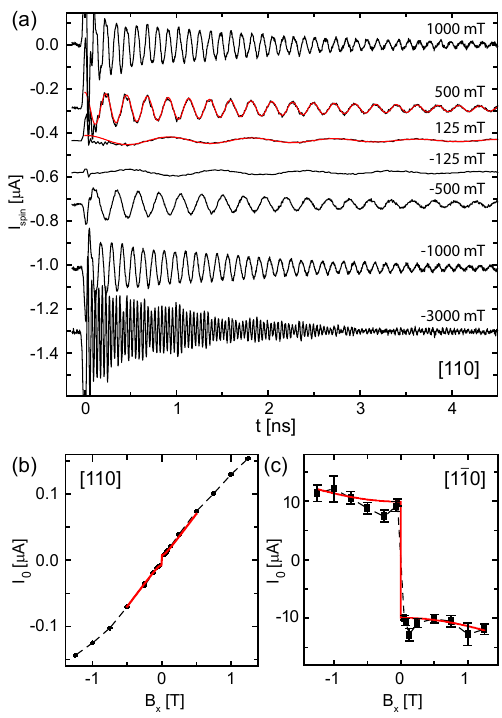}
\caption{ \label{fig2} $B$-field dependence of spin precession driven electric current. (a), Time-resolved current along the $[1 1 0]$ direction in InGaAs at various $B$ fields. Larmor precession frequency and magnitude of AC current increase with increasing $B$ field strength while the sign of the AC current reverses when reversing the $B$ field direction. (b) Amplitude $I_0$ of AC current vs $B_x$. The linear increase of $I_0$ with $B_x$ indicates a strong PDTM with a weak SGE visible at small $B_x$. The red solid line is a fit to Eq.~\eqref{Separation1}. (c) $I_0$ vs  $B_x$ for a sample measured along the $[1\overline{1}0]$ direction. The weak increases of $I_0$ with increasing $|B_x|$ indicate the dominance of SGE over PDTM.}
\end{figure}

To further explore the dependence of the spin-driven AC current on $B_{x}$, we show a series of measurements at selected $B$ fields in \fg{fig2}(a).
It is apparent that both the frequency $\omega_L$ and the amplitude $I_0$ increase with increasing $B$ field strength (see also \fg{fig2}(b)). Starting from $B_x=0.05$\,T, when the oscillations become visible, the current frequency scales linearly with $B_x$ and coincides with the spin precession frequency seen in TRFR measurements. At large $B_{x} = \pm3$\,T the spin precession frequency reaches $f=26$\,GHz (see \fg{fig2}(a)). Remarkably, the amplitude $I_0$ of the AC current also increases linearly with $B_{x}$ (deviations at small and at large $B$ fields are discussed further below). This is in contrast to the spin dynamics studied by TRFR, where the signal amplitude is a measure of the total spin density which is independent of $B$, i.e. the spin precession frequency. When reversing the $B$ field direction, $I_0$ changes sign indicating that the AC current generation depends on the spin precession direction which changes from clockwise to counterclockwise upon $B$ field reversal.

All these findings suggest that the precessing spin polarization drives the AC charge current, i.e. it triggers a periodic trembling motion of the electrons. A direct coupling between the electron spin and velocity is indeed possible in semiconductor structures with ${\bm k}$-linear SOI~\cite{IvchenkoGanichev2017}.
%~\cite{Kato2004_manipulation}. 
Here, we use $n$-InGaAs epilayers which
exhibit both the Dresselhaus SOI due to strain by lattice mismatch to the subjacent Si-GaAs substrate and the Rashba-like SOI due to partial strain relaxation~\cite{PhysRevB.72.115204}. 
Semi-classical theory~\cite{Ivchenko_JETP_1990} suggests that a pulsed optical excitation 
by circularly polarized light in an external in-plane magnetic field may lead to spin-related charge current oscillations. The theory is based on the effective electron Hamiltonian
\begin{equation}
H = \frac{\hbar^2k^2}{2m^*} + \frac{\hbar}{2} {\bm \omega_L} \cdot {\bm \sigma} +  \beta_{xy} \sigma_{x} k_{y} + \beta_{yx} \sigma_{y} k_{x}\:,
\label{OurHamiltonian}
\end{equation}
where $m^*$ is the effective mass, $\beta_{xy}$ and $\beta_{yx}$ are the constants originating from Rashba/Dresselhaus SOI in (001)-oriented structures, $\sigma_i$ are the Pauli spin matrices, 
and $x$ and $y$ are the in-plane axes. 
The current contains two contributions. The first one emerges due to the spin relaxation of polarized electrons and is governed by the relaxation rate ${\bm S}(t)/T_2^*$. This effect, known as the spin-galvanic effect (SGE), was observed and studied experimentally under steady-state photoexcitation in GaAs-based quantum wells~\cite{Nature417_Ganichev2002_Spin-GalvanicEffect,PhysRevB.68.081302}. 
The second contribution can be referred to as precession driven trembling motion (PDTM). 
%The second contribution 
emerges due to the dynamic precession of the spin polarization vector in a magnetic field and is governed by the vector product ${\bm \omega}_L \times {\bm S}(t)$.      
In (001)-oriented structures, the SGE current is driven by the \textit{in-plane} spin component, i.e. it changes sign when reversing the in-plane spin direction from the $+y$ to $-y$ orientation (see \fg{fig1}(f)) and its amplitude $I_{\rm SGE}$ does not depend on $B_{x}$~\cite{Nature417_Ganichev2002_Spin-GalvanicEffect,PhysRevB.68.081302}.
As the AC charge current in \fg{fig1}(d) becomes largest when electron spins are oriented in the out-of-plane direction along the $z$-axis and its amplitude linearly increases with $B_x$, it cannot result from the SGE. We note, however, that we also observe the SGE in our time-resolved AC charge current measurements. This is most clearly seen along the $[1\overline{1}0]$ direction where we observe the expected phase difference of ($\phi \sim \pi/2$) between the respective Larmor precessions in the AC charge current in \fg{fig1}(g) and the TRFR in \fg{fig1}(h) proving the AC charge current is generated by in-plane spins. As expected for the SGE, the respective current amplitudes are roughly independent on $|B_{x}|$ and change sign when reversing the magnetic field direction (\fg{fig2}(c)). 

Because of the $\pi/2$ phase difference between the SGE and PDTM currents their amplitudes, $I_{\rm SGE}$ and $I_{\rm PDTM} = K_{\rm PDTM} B_{x}$, respectively, add quadratically to the AC current amplitude
\begin{equation}
I_0(B_x)=\sqrt{I_{\rm SGE}^2+(K_{\rm PDTM} B_{x})^2}, \label{Separation1}\\
\end{equation}
where $K_{\rm PDTM}$ is the strength of PDTM. This equation describes well the data measured along both crystal directions shown in Fig.~2. While the PDTM dominates over the SGE for the $[110]$ sample Figs.~2(b) it is the opposite for the $[1\overline{1}0]$ sample in Fig.~2(c). The slight deviation from the expected linear behavior at large magnetic fields in Fig.~2(b), i.e. at large precession frequencies is due to the limited band-width of our device (3db point at 13 GHz).

\begin{figure}[tbp]
\includegraphics[width=0.6\linewidth]{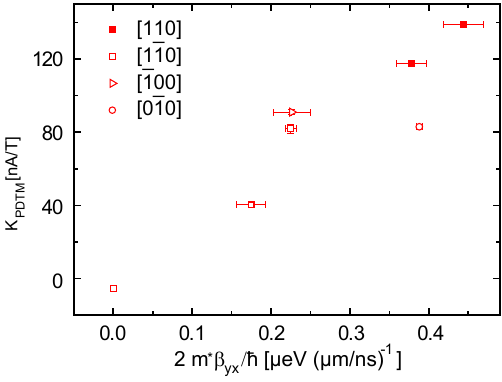}
\caption{\label{fig3} Dependence of coupling strength $K_{\rm PDTM}$ of the periodic trembling motion on the spin-orbit coupling constant $\beta_{yx}$. The data points are obtained from a number of InGaAs devices where the AC current is probed along different crystal directions. The overall linear behavior is suggesting that SO coupling is decisive to the generation of the PDTM component of the AC current.}
\end{figure}

We next explore how $K_{\rm PDTM}$ depends on the SOI. The crystal axis anisotropy of the SOI gives us a set of devices from the same wafer with different $\beta_{yx}$ (Supplemental Material \cite{supp}). For measuring $\beta_{yx}$ we use the TRFR pump-probe technique and apply an additional DC current $I$ which converts into an internal magnetic field $B_{\rm int}\propto I$ (Refs.~\onlinecite{Meier_Rashba_2007, Kato2004_manipulation}). From the change in the spin precession frequency we determine $\beta_{yx}$ for each device and find both Rashba and Dresselhaus contributions leading to highly anisotropic SO fields. We conducted measurements on seven devices along four crystal directions (see \fg{fig1}a) and extracted the respective $K_{\rm PDTM}$ and $\beta_{yx}$ values which are summarized in \fg{fig3}. Despite a certain device-to-device variation along the same crystal direction, we clearly observe an overall linear dependence of $K_{\rm PDTM}$ on $\beta_{yx}$ demonstrating that the SO coupling plays the key role for the emergence of the PDTM.

While the AC current % driven by the in-plane spin polarization can be well understood by the SGE, the current 
dynamically generated by electron spin precession can be understood semiclassically as outlined above, we now present a quantum-mechanical picture as a result of coherent electron Zitterbewegung triggered by optical pulses~\cite{TarasenkoJETPL2018} which results in a trembling motion of electrons at the Larmor frequency of the precessing electron spin ensemble. This interpretation originates from the fact that the electron velocity is not a conserved quantity in the presence of the Zeeman gap and spin-orbit interaction. 
Indeed, besides the first term of spin-independent kinetic energy, the effective Hamiltonian~\eqref{OurHamiltonian}
mimics the Dirac Hamiltonian with all essential elements: it comprises the SOI term which provides a $\bm k$-linear coupling of the states and the Zeeman term which opens a gap at $\bm k=0$ and plays a role of the mass term in the Dirac Hamiltonian.

Calculating the electron velocity operator $v_x = (i/\hbar) [H,x]$, where $[A,B]=AB-BA$ is the commutator of the operators $A$ and $B$,
and then the acceleration operator $\dot{v}_x = (i/\hbar) [H,v_x]$, we obtain to first order in SOI
\begin{equation}
\dot{v}_x = -\frac{\beta_{yx}}{\hbar}\omega_L\sigma_z \:.
\label{velocity_operator}
\end{equation}

In our InGaAs samples, the electron transport is diffusive rather than ballistic and we deal with an ensemble of about $10^8$ electrons.
To describe the macroscopic motion of the electrons we average Eq.~\eqref{velocity_operator} over the electron ensemble. Since Eq.~\eqref{velocity_operator} is linear, this leads to the replacement  of the operators ${v}_x$ and ${s}_z$ by their average values $\bar{v}_x$ and $\bar{s}_z$. Besides, to account for electron scattering by phonons or static defects, which randomizes electron trajectories and  slows down the average electron velocity, we add a Drude-like term and obtain
\begin{eqnarray}
\dot{\bar{v}}_x(t) = -2\frac{\beta_{yx}}{\hbar}\omega_L \bar{s}_z(t) - \frac{\bar{v}_x(t)}{\tau_p} \:,
\label{velocity}
\end{eqnarray}
where $\tau_p$ is the momentum scattering time. It is assumed that $T_2^*\gg 1/\omega_L, \tau_p$ with $T_2^*$ being the spin dephasing time.

The AC charge current density caused by the PDTM of the electrons is given by $j^{(ac)}_x(t) = e n_e \bar{v}_x(t)$, where $e$ is the electron charge and $n_e$ is the electron density. For the case $\omega_L\tau_p\ll 1$, which  is realized in our experiments, the solution of Eq.~\eqref{velocity} yields
\begin{eqnarray}
j^{(ac)}_x(t) = -2en_e\frac{\beta_{yx}}{\hbar} \omega_L \tau_p \bar{s}_z(t) \:.
\label{current_density}
\end{eqnarray}
As it follows from Eq.~\eqref{current_density}, the amplitude of the PDTM current which is determined by $K_{\rm PDTM}$, scales linearly with the SO coupling constant $\beta_{yx}$ (see \fg{fig3}). We note that the coupling of the macroscopic current density to the average spin given by Eq.~\eqref{current_density} was previously derived by a sophisticated spin-density-matrix approach in Ref.~\cite{Ivchenko_JETP_1990}. Here, we have shown the quantum-mechanical derivation of this result and uncover its relation to the physics of Zitterbewegung explaining our experimental findings.

For illustration, we depict in \fg{fig1}(b) the temporal evolution of the spin vector $\bar{s}(t)$ during spin precession and the predicted AC charge current according to Eq.~\eqref{current_density}. Spin precession and trembling movement are in phase, i.e., the largest current along the $+x$ direction is obtained when the spin is pointing along $+z$ (red arrow). The current becomes zero at the spin rotation angle $\pi/2$ ($S_z=0$, see yellow arrow in \fg{fig1}(b)), changes the sign for larger angles, and again becomes strongest along the $-x$ direction for the angle $\pi$ (green arrow). The PDMT current exhibits one distinct frequency which linearly increases with increasing magnetic field strength, despite the randomized electron momenta present in the diffusive transport regime. For $B\neq0$, the spin projection $\bar{s}_z(t)$ oscillates at the Larmor frequency $\omega_L$ and so does the electron velocity -- which results in the AC charge current -- even though no driving electric field is applied.
We note that the current amplitude estimated from Eq.~\eqref{current_density} for the sample parameters determined independently (see Supplemental Material \cite{supp}) $I_0 \sim 0.2$~$\mu$A is in agreement with the experimental data.

We have shown that electrons  in III-V semiconductors with spin-orbit coupling experience an inherent trembling motion of quantum-mechanical nature. The trembling motion (Zitterbewegung) of individual electrons can be phase-synchronized by initializing the electrons in the same spin states and detected as an AC current oscillating at the Larmor frequency of several GHz. The use of pulsed optical excitation combined with time-resolved electrical detection in systems with rather small Zeeman splitting provides a promising pathway for exploring trembling motion in other solid state systems. Additionally, the spin-driven AC currents can be utilized as an ultrafast spin sensitive electrical probe of electron spin precession where the detection scheme is solely based on SO interaction free of any additional spin-sensitive magnetic materials. We expect that spin-orbit-driven high frequency currents can also be explored in other materials including topological insulators with inherent Dirac-like spectrum and other 2D materials with strong spin-orbit interaction and short spin lifetimes.

\section{Methods}
\subsection{Sample fabrication}

An n-doped In$_{0.07}$Ga$_{0.93}$As epilayer was grown via molecular beam epitaxy on a semi-insulating (001) GaAs substrate, with a doping concentration of 8.25~$\times$~10$^{16}$~cm$^{-2}$. The epilayer was lithographically patterned into pentagonal structures, and ohmic contacts were formed by depositing Ge/Au/Ni stacks, followed by rapid thermal annealing at 400~$^\circ$C.

\subsection{High-frequency measurement setup}
For broadband electrical measurements, the samples were mounted on custom-designed printed circuit boards (PCBs) incorporating 50~$\Omega$ coplanar waveguides (CPWs). The CPWs were fabricated on RO3210 laminate using UV lithography and wet etching. Electrical coupling between the InGaAs channels and CPWs was achieved using three short Al bonding wires. The CPWs were connected to high-frequency coaxial cables via impedance-matched SMA launchers and linked to a 26~GHz bandwidth amplifier with 35~dB gain. 

\subsection{Optical excitation and detection}

Phase-coherent spin ensembles in the InGaAs layer were excited using circularly polarized 3~ps laser pulses generated by a mode-locked Ti:sapphire laser (\textsc{Spectra Physics} Tsunami) operating at a repetition rate of 80~MHz. The polarization of the pump beam was switched between $\sigma^+$ and $\sigma^-$ using a liquid crystal variable retarder. For time-resolved electrical detection we recorded the electrical signal by a fast sampling oscilloscope (Tektronix DSA 8200), phase-locked to the laser pulses via a fast photodiode, ensuring precise temporal resolution of the GHz-range oscillatory voltage signals. 

For time-resolved Faraday rotation (TRFR), a second, linearly polarized probe beam was used to monitor spin precession. The TRFR signal was detected with a balanced photodiode bridge and demodulated using lock-in amplifiers for enhanced sensitivity. For details of the optical setup see Section~6 of the Supporting Information of Ref.~\cite{Ersfeld2020May}. All measurements were conducted in a helium bath cryostat (\textsc{Oxford}, Spectromag) with magnetic fields applied along the in-plane sample axis.

\section{Data availability}
The data that support the findings within this paper are available at 10.5281/zenodo.15968129

\begin{acknowledgments}
We thank G. G\"untherodt and F. Volmer for helpful discussions, S. Pissinger for initial work on time-resolved photo-current measurements, and S. Staacks for help on the figures. This work was supported by the Deutsche Forschungsgemeinschaft (DFG, German Research Foundation) under Germany's Excellence Strategy - Cluster of Excellence Matter and Light for Quantum Computing (ML4Q) EXC 2004/1 – 390534769 and by RSF grant 22-12-00211-$\Pi$.
\end{acknowledgments}


\begin{thebibliography}{36}
\bibitem [{sup()}]{supp}%
  \BibitemOpen
  \href@noop {} {}\bibinfo {howpublished} {See Supplemetal Material [URL will be inserted by publisher] for additional details on the spin galvanic effect, the absence of the periodic trembling motion in static experiments and the mapping of SOI fields.}\BibitemShut {Stop}%
\end{thebibliography}
\end{document}

%___________________________________________________________________________________________
%
% title
%___________________________________________________________________________________________

\title{Supplementary Information: Trembling Motion of Spin Polarized Electrons}

\author{I. Stepanov}
\affiliation{2nd Institute of Physics and JARA-FIT, RWTH Aachen University, D-52074 Aachen, Germany}

\author{M. Ersfeld}
\affiliation{2nd Institute of Physics and JARA-FIT, RWTH Aachen University, D-52074 Aachen, Germany}

\author{A. V. Poshakinskiy}
\affiliation{Ioffe Institute, 194021 St Petersburg, Russia}

\author{M. Lepsa}
\affiliation{Peter Gr\"{u}nberg Institut (PGI-9) and JARA-FIT, Forschungszentrum J\"{u}lich GmbH, D-52425 J\"{u}lich, Germany}

\author{E. L. Ivchenko}
\affiliation{Ioffe Institute, 194021 St Petersburg, Russia}

\author{S. A. Tarasenko}
\affiliation{Ioffe Institute, 194021 St Petersburg, Russia}

\author{B. Beschoten}
\affiliation{2nd Institute of Physics and JARA-FIT, RWTH Aachen University, D-52074 Aachen, Germany}

\maketitle

\subsection*{A. Influence of the spin-galvanic effect}

The spin dependent current oscillating at the Larmor frequency arises generally from two components: the precession driven trembling motion of spin polarized electrons (PDTM) and the spin-galvanic effect caused (SGE) caused by spin relaxation of polarized electrons. In most cases, the signal is dominated by PDTM. However, SGE is clearly present in many measurements and can be dominant under certain conditions.

This subsection presents a detailed description of the difference between PDTM and SGE in our experiment. PDTM and SGE are driven by different microscopic mechanisms resulting in distinctly different behaviors of the spin dependent current in time-resolved measurements. PDTM is a spin precession-driven effect and does not involve spin flip scattering \cite{Zawadzki_ZitterbewegungReview}. As shown in the main text, the PDTM signal is proportional to the precession frequency $\omega_L\propto B_{x}$ and to the out-of-plane spin polarization $S_z(t)$ and vanishes at $B=0$~T. The SGE, on the other hand, is driven by the asymmetry of the spin-flip scattering in the presence of SO splitting \cite{Ivchenko_JETP_Letters_1989,Ivchenko_JETP_1990,Nature417_Ganichev2002_Spin-GalvanicEffect,JETPL85_Golub2007_NewMechanismoftheSpin-GalvanicEffect}. The current generated by SGE is directly proportional to the total in-plane spin polarization $S_y(t)$ and does therefore not change with the magnetic field strength.

The initial spin polarization is phase-triggered by circularly polarized laser pulses with $S_z(t=0)=max$. Accordingly, we can easily distinguish SGE from PDTM by their time dependent signatures:
\begin{gather}
I_{PDTM}(t)=K_{PDTM}\cdot B_{x} \cdot\cos(\omega_Lt)\cdot\exp\left(-\frac{t}{\tstern}\right), \label{Component2}\\
I_{SGE}(t)=I_{SGE}\cdot\sin(\omega_Lt)\cdot\exp\left(-\frac{t}{\tstern}\right), \label{Component1}
\end{gather}
where $K_{PDTM}$ accounts for the strength of the PDTM oscillating current and $I_{SGE}$ is the amplitude of the SGE current. The strength of PDTM and of SGE vary from sample to sample and particularly for PDTM it depends on the crystal direction and the strength of spin-orbit interaction. \fg{supp_fig1} shows two spin current measurements taken along different crystal directions at $B_{x}=125$~mT. Spin precession with the same Larmor frequency is observed in both cases. We note that spin precession at $t=0$~ns starts with different phases. The spin current from PDTM is largest at $t=0$~ns following a cosine function (Eq.\eqref{Component2}), see also main text) while the spin current from SGE is zero at $t=0$~ns and becomes largest after $\pi/2$ spin precession into the in-plane $y$-direction following a sine function (Eq.\eqref{Component1}). The SOI along $[1\bar1 0]$ is relatively weak, leading to a small amplitude of PDTM and to a pronounced SGE. The SOI along $[1 1 0]$, on the other hand, is very strong with a very strong PDTM, which is dominating the signal on this sample.

\begin{figure}[h]
\includegraphics{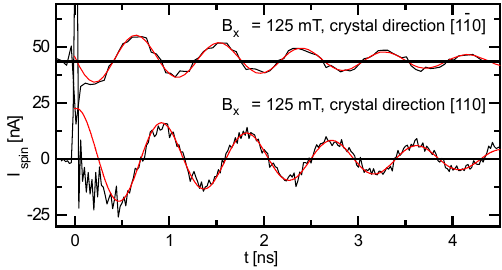}
\caption{\label{supp_fig1} Spin dependent AC current at $B_{x}=125$~mT for crystal directions $[1\bar1 0]$ and $[1 1 0]$ (with an vertical offset for clarity). The curves show two extreme examples, with the signal along $[1\bar1 0]$ dominated by SGE and the signal along $[1 1 0]$ dominated by PDTM. The fits to Eq. \eqref{Oszillation} reveal the difference in phase $\phi$ due to different effects. The curve along $[1 \bar1 0]$ is mostly a sine curve with $\phi=88.8^\circ$, which is close to $\phi=\pm 90^\circ$ typical for SGE. The curve along $[1 1 0]$ is mostly a cosine curve with $\phi=-10.3^\circ$, which is close to $\phi=0^\circ$ typical for PDTM.}
\end{figure}

Usually, we observe a superposition of both effects. The resulting AC current signal can then be fitted to
\begin{equation}
I_{spin}(t)=I_{PDTM}(t)+I_{SGE}(t)=I_0\cdot\exp\left(-\frac{t}{\tstern}\right)\cdot\cos\left(\omega_L t+\phi\right),
\label{Oszillation}
\end{equation}
where $I_0$ is the amplitude and $\phi$ is the phase of the resulting AC current. Because of the $90^\circ$ phase difference between SGE and PDTM, both amplitudes add quadratically and the phase of their superposition $\phi$ can be calculated by a simple trigonometric function:
\begin{gather}
I_0=\sqrt{I_{SGE}^2+(K_{PDTM}\cdot B_{ext})^2}, \label{Separation1}\\
\cot{\phi}=-\frac{K_{PDTM}\cdot B_{ext}}{I_{SGE}}. \label{Separation2}
\end{gather}
The measured dependence of the amplitude $I_0$ and the phase $\phi$ on $B_{x}$ from a fit to Eq. \eqref{Oszillation} is shown in \fg{supp_fig2}.

\begin{figure}[h]
\includegraphics[width=0.7\textwidth]{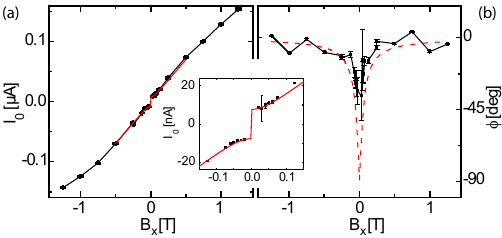}
\caption{\label{supp_fig2} Dependence of amplitude $I_0$ and phase $\phi$ on $B_{x}$ for crystal direction $[1 1 0]$ \textbf{a}, $I_0(B_{x})$, showing a very strong PDTM. Red curve is a fit to Eq. \eqref{Separation1} with $I_{SGE}=9.2$~nA and $K_{PDTM}=139$~nA/T. \textbf{b} dependence of $\phi$ on $B_{x}$. The dashed curve shows the prediction of the phase using $I_{SGE}$ and $K_{PDTM}$ extracted from the fit of the amplitude. The data and the prediction agree well with a strong PDTM as $\phi$ is close to 0$^\circ$ for a wide range of $B_{x}$.}
\end{figure}

It is straightforward to extract $I_{SGE}$ and $K_{PDTM}$ from the dependence of $I_0$ on $B_{x}$ in \fg{supp_fig2} a. The data shows that $I_0$ from measurements along $[1 1 0]$ increases linearly with $B_{x}$ and shows a $B_{x}$-independent contribution from SGE only at very small magnetic fields as seen in the inset of \fg{supp_fig2} a. This indicates a very strong PDTM and rather small SGE signal which is only unveiled at small magnetic fields. Although the AC current from SGE are generated at all magnetic fields and its magnitude is independent on the magnetic field strength, it only becomes dominant at very low magnetic fields when the PDTM contribution approaches zero. Using the values of $I_{SGE}$ and $K_{PDTM}$ we can predict the dependence of $\phi$ on $B_{x}$, shown as a dashed line in \fg{supp_fig2} b, and compare it to the data. We see a good agreement with the data, showing $\phi$ values close to $0^\circ$ and typical for PDTM for a wide range of $B_{x}$.

\begin{figure}[h]
\includegraphics[width=0.7\textwidth]{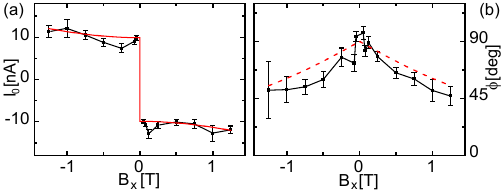}
\caption{\label{supp_fig3} Dependence of amplitude $I_0$ and phase $\phi$ on $B_{x}$ for crystal direction $[1 \bar{1} 0]$ \textbf{a}, $I_0(B_{x})$, showing a very weak PDTM. The signal is dominated by SGE. Red curve is a fit to Eq. \eqref{Separation1} with $I_{SGE}=9.8$~nA and $K_{PDTM}=-5.4$~nA/T. \textbf{b} dependence of $\phi$ on $B_{x}$. The dashed curve shows the prediction of the amplitude using $I_{SGE}$ and $K_{PDTM}$ extracted from the fit of the amplitude. The data and the prediction agree well with a weak PDTM as $\phi$ is close to 90$^\circ$ for a wide range of $B_{x}$ and is only slowly shifting towards 0$^\circ$.}
\end{figure}

\fg{supp_fig3} shows the magnetic field-dependence of spin-dependent currents along the $[1\bar1 0]$ crystal direction. In contrast to the crystal direction $[1 1 0]$ there is only a small increase in amplitude with $B_x$. Fitting the data reveals SGE contributions that are almost identical along both crystal directions. In contrast, the PDTM contributions becomes vanishingly small and negative along the $[1\bar1 0]$ direction.

The reason for the different PDTM strength between crystal directions $[1 1 0]$ and $[1\bar1 0]$ is the different spin-orbit coupling strength. For currents flowing along $[1 1 0]$, Rashba and Dresselhaus SOI have the same orientation, resulting in a strong PDTM. On the other hand, for currents flowing along the $[1\bar1 0]$ crystal direction, Rashba and Dresselhaus field point in opposite directions and add up to a small total values of SO coupling strength resulting in a vanishingly small PDTM.

\subsection*{B. SGE and PDTM in static experiments}
As explained above, both PDTM and SGE are relying on different microscopic mechanisms and have distinctly different signatures in time-resolved measurements. In this section we present calculations and experimental results showing that despite these obvious differences, SGE and PDTM are indistinguishable in static experiments, making time-resolution essential for their observation and analysis.

Because the SGE voltage is generated by an in-plane spin polarization, its time-evolution after optical excitation is described by an exponentially decaying sine function. As is was shown in \cite{Nature417_Ganichev2002_Spin-GalvanicEffect}, the time-integration of the SGE voltage results in an antisymmetric Hanle curve
\begin{equation}
 \bar{U}_{SGE}(\omega_L)\propto \int_{0}^\infty \sin(\omega_L t) \mathrm e^{-t/ \tstern}\,dt=
\frac{\omega_L {\tstern}^2}{1+(\omega_L\tstern)^2},
\label{HanleFit}
\end{equation}
where $\tstern$ is the spin dephasing time and $\omega_L$ is the Larmor frequency proportional to the applied magnetic field $B_{x}$. This antisymmetric Hanle curve describes the magnetic field dependence of the detected voltage during continuous optical excitation or after pulsed optical excitation when using time-averaging voltage measurement techniques.

The voltage generated by PDTM, on the other hand, is proportional to the out-of-plane spin polarization $S_z$ and to the precession frequency $\omega_L$. For a signal proportional to $S_z$, which follows a cosine function, we expect a symmetric Hanle curve. However, when also considering the frequency dependent PDTM amplitude in Eq. \eqref{Component2} with $\omega_L\propto B_{x}$ it can be shown that the time-integral $\bar{U}_{PDTM}(B_{x})$ is again an antisymmetric Hanle curve:
\begin{eqnarray}
 \bar{U}_{PDTM}(\omega_L)\propto \omega_L \cdot\int_{0}^\infty  \cos(\omega_L t) \mathrm e^{-t/ \tstern}\,dt=\\
\omega_L\cdot\frac{\tstern}{1+(\omega_L\tstern)^2}.
\end{eqnarray}
Both effects are thus expected to have an identical signatures in static experiments although they behave distinctly different in time-resolved measurements.
\begin{figure}[h]
\includegraphics[width=0.5\textwidth]{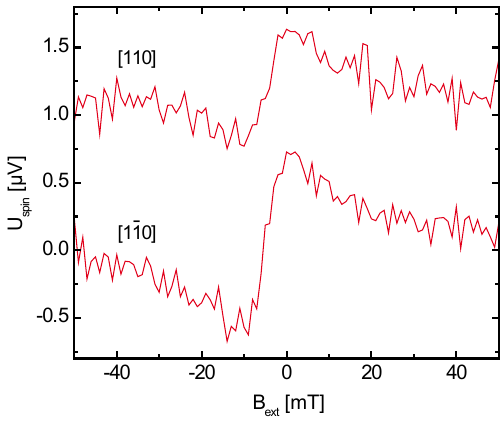}
\caption{\label{supp_fig4} Static measurements of spin-dependent voltages along $[1\bar1 0]$ dominated by SGE and along $[1 1 0]$ dominated by PDTM with a vertical offset for clarity. Both measurements are identical in their signature and do not allow to distinguish their underlaying generation mechanism.}
\end{figure}
To prove this notion, we conducted static measurements of spin photo-voltages. We used pulsed laser excitation with an average power on 30 mW and intensity modulation at 400 Hz to allow for lock-in detection of the resulting voltage averaging the signal over 2 seconds. As for the time-resolved experiments, the spin-dependent signal was extracted by subtracting the photo-voltages generated by $\sigma^+$ and $\sigma^-$ excitation: $U_{spin}=(U(\sigma^+)-U(\sigma^-))/2$. The results of the measurement along the crystal axes $[1\bar1 0]$ and $[1 1 0]$, are shown in \fg{supp_fig4}. Although the spin voltages are dominated by SGE and by PDTM along $[1\bar1 0]$ and $[1 1 0]$, respectively, both measurements show antisymmetric Hanle curves as expected from the above calculations. This finding demonstrates that time-resolution measurements are of utmost importance to unveil the underlying spin dynamics.
\newpage
\subsection*{C. Mapping of SOI field}
The InGaAs layers exhibit spin-orbit interactions with both Rashba and Dresselhaus fields. The strain-induced Dresselhaus field is due to lattice mismatch between the InGaAs layer and the GaAs substrate, while Rashba field is due to strain gradient in the growth direction\cite{PhysRevB.72.115204,Orders1987}. To map the SOI fields along different crystal directions we adapt a method similar to \cite{Meier_Rashba_2007}. We use a standard double modulated pump-probe time-resolved Faraday rotation measurement technique to create and to probe coherent electron spin ensembles. Their spins precess about the total magnetic field $B_{tot} = B_{ext}+B_{int}$, which is a combination of the external magnetic field $B_{ext}$ and the internal magnetic field $B_{int}$, created by the applied current density $j$ in the presence of SOI \cite{Meier_Rashba_2007,Kato2004_manipulation}. \fg{supp_fig5}~a shows the influence of $B_{int}$ on spin precession at  $B_{ext}=50$\,mT for three different current densities measured at $T=50$\,K. Because $B_{int}\perp B_{ext}$ for $j\parallel[110]$, $B_{tot}$ is always increasing with $j$ leading to faster spin precession.

We can separate the components of $B_{int}$ parallel $B_{\parallel}=c_{\parallel}\cdot j$ and perpendicular $B_{\perp}=c_{\perp}\cdot j$ to $B_{ext}$ which are both proportional to the current density $j$. Vector addition leads to the expression

\begin{equation}
 \mid B_{tot}\mid (j)=\sqrt{(c_{\perp}\cdot j)^2+(B_{ext}+c_{\parallel}\cdot j)^2}.
\label{bint-separation}
\end{equation}

\fg{supp_fig5}~b shows the values of $\mid B_{tot}\mid (j)$ for a set of $B_{ext}$ along $[1 1 0]$ which we determine from TRFR measurements with $g=0.621$. We fit the data to the expression \eqref{bint-separation} and observe a parabolic curve of $\mid B_{tot}\mid (j)$, resulting from $B_{int}\perp B_{ext}$ without any parallel component. This behavior is expected along $[1 1 0]$ because Rashba and Dresselhaus fields only have perpendicular components of $B_{int}$ for $k$ along this crystal direction. \fg{supp_fig5}~c, on the other hand, shows an example with $B_{ext}$ and $j$ along $[0\bar1 0]$ exhibiting components $B_{int}\perp j$ originating from the Rashba field and $B_{int}\parallel j$ originating from the Dresselhaus field, evident as a slope in $\mid B_{tot}\mid (j)$ in addition to the parabolic shape. We determine the sign of $B_{int}\perp j$ by repeating the measurement along the same crystal direction but on another sample rotated by $90^\circ$. This allows determining the SOI strength for every individual sample.

Knowing the charge carrier density $n_e=8.25\cdot 10^{16}$\,cm$^{-3}$ at $T=50$\,K from Hall effect measurements, we use the Drude model to calculate the drift velocity from the current density $j$. With the g-factor $g=0.621$ we can calculate the energy splitting due to SOI from the measured $B_{int}$. The strength and direction of SOI for samples used in this study are mapped in \fg{supp_fig5}~d. Each arrow represents one individual sample. For electron drift along $[1 1 0]$ and $[1\bar1 0]$ crystal directions we observe mostly perpendicular spin splitting due to Rashba and Dresselhaus SOI (constructive superposition for currents along $[1 1 0]$). For electron drift along $[\bar1 0 0]$ and $[0\bar1 0]$, we observe components of spin splitting parallel and perpendicular do drift direction, with a sign change of the parallel component as a result of Dresselhaus SOI.

\begin{figure}[h]
\includegraphics{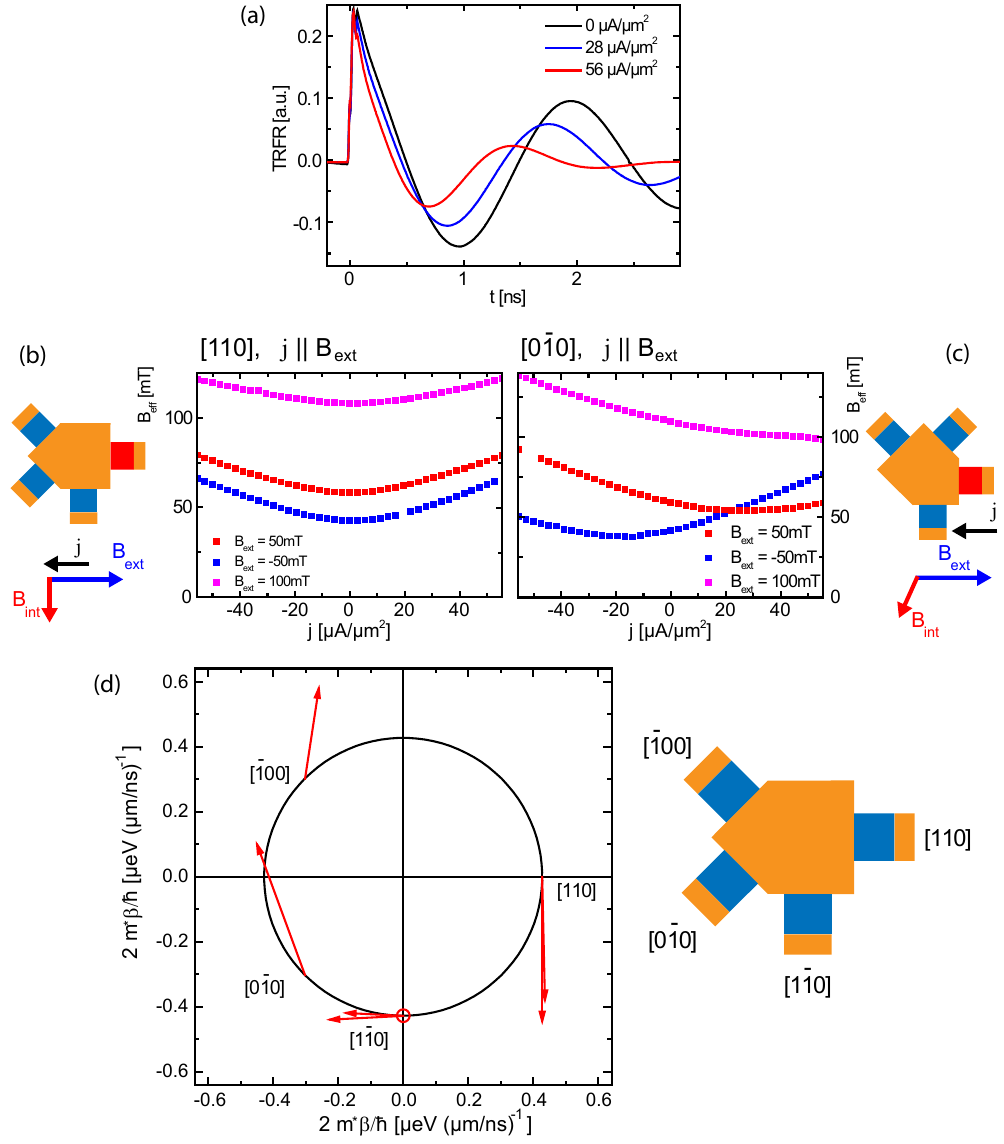}
\caption{\label{supp_fig5} \textbf{a}, TRFR measurements of spin precession under the influence of SOI at three different current densities. \textbf{b}, effective magnetic field in dependence of applied current along $[1 1 0]$ with $B_{ext}$ of 50\,mT, -50\,mT and 100\,mT. The schematics left shows the orientation of the sample, $B_{ext}$ and the extracted $B_{int}$, which is in this case perpendicular to $j$. \textbf{c}, effective magnetic field for $j$ along $[0\bar1 0]$ revealing components parallel and perpendicular to $j$. \textbf{d}, map showing strength and direction of spin-orbit splitting for a variety of samples, revealing a superposition of Rashba and Dresselhaus SOI. The open circle along $[1\bar1 0]$ represents an additional sample with $\beta=0$.}
\end{figure}

\bibliographystyle{apsrev}
\selectlanguage{english}
\bibliography{literature}